# Determination of the levitation limits of dust particles within the sheath in complex plasma experiments

Angela Douglass, Victor Land, Ke Qiao, Lorin Matthews, and Truell Hyde[a]

*Center for Astrophysics, Space Physics, and Engineering Research,*
*Baylor University, Waco, TX, USA 76798-7316, www.baylor.edu/CASPER*

Experiments are performed in which dust particles are levitated at varying heights above the powered electrode in a RF plasma discharge by changing the discharge power. The trajectories of particles dropped from the top of the discharge chamber are used to reconstruct the vertical electric force acting on the particles. The resulting data, together with the results from a self-consistent fluid model, are used to determine the lower levitation limit for dust particles in the discharge and the approximate height above the lower electrode where quasineutrality is attained, locating the sheath edge. These results are then compared with current sheath models. It is also shown that particles levitated within a few electron Debye lengths of the sheath edge are located outside the linearly increasing portion of the electric field.

---------
a) Electronic mail: Truell_Hyde@baylor.edu

Determination of the levitation limits of dust particles within the sheath in complex plasma experiments

**I. INTRODUCTION**

Complex plasma physics is the study of partially ionized gases containing small, usually micron-sized, dust particles. These particles collect ions and electrons from the plasma and obtain a negative charge due to the high mobility of the electrons (typically 3 elementary charges per nanometer radius[1]). Consequently, on Earth, they are trapped within a non-neutral region in the plasma that forms in front of the lower electrode, often referred to as a sheath. The strong electric fields that form in these regions can levitate dust particles against the force of gravity. The levitation of the particles therefore depends on the electric field profile within the sheath, as well as on the charge-to-mass ratio of the particles.

A complete understanding of the levitation of dust particles within the sheath region is a complicated matter for two reasons: first of all, a complete and self-consistent theory for this non-neutral region (sheath) is non-existent. Despite many solutions for the vertical dependence of the electric field at the height where dust particles are levitated (usually found to be linearly dependent on $z$[2]), the validity of these solutions are still heavily debated[3,4]. Secondly, the charging of dust particles within the sheath is a complicated problem. Calculation of the dust charge must include modification of the commonly assumed Maxwellian distributions, particularly for ions, as well as corrections for the deviation from quasineutrality due to electron depletion[5]. Both the electric field and the dust charge could be calculated if the plasma parameters within the sheath could be experimentally determined, but direct measurements of the plasma parameters within the sheath are very hard to obtain. Probes cannot accurately measure the plasma parameters within this region since they have been found to alter the plasma parameters. Therefore, a seemingly simple question: "What are the plasma characteristics at the levitation height of the dust particles in complex plasma experiments?" cannot be immediately answered in many cases.

In this paper, two experiments in which dust particles are used as probes for the sheath are presented. The experimental results are compared with numerical results to locate the approximate height above the lower electrode where quasineutrallity is attained, locating the sheath edge, and the extent of the unstable region near the lower electrode, locating the lower levitation limit for dust particles, in Radio Frequency (RF) discharges. These results are then compared to current sheath theories. With this, a new, *in situ* experimental method is provided that can be employed in any dusty plasma system to determine the approximate height of the

Determination of the levitation limits of dust particles within the sheath in complex plasma experiments sheath edge. It will also be shown that the electric field experienced by particles located near this boundary cannot be described by the classical, linear theory which assumes an electric field of zero or $-k_B T_e / e\lambda$ at the sheath edge[6] and a magnitude that decreases linearly with increasing height across the entire sheath. An approximately linear electric field is found in the central section of the sheath, but deviations from this linear segment are found at both the upper and lower edges of the sheath. Finally, the lower levitation limit is found through the use of a stability criterion derived from the force balance equations.

First, the background information will be presented, including a short overview of current sheath theories, a short description of dust particle charging and a derivation of the lower instability criterion for dust particle levitation. In Section III, a description of the experiments performed will be provided, as well as a brief description of the numerical model employed. Results from the experiments and the fluid model are presented in Section IV. Finally, a discussion of the results and conclusions drawn from these results are given in Sections V and VI, respectively.

## II. BACKGROUND

### A. The classical sheath

A non-neutral region called a sheath forms in front of plasma-facing surfaces to balance the electron and ion losses to the surface. In one of the first treatments of the sheath[7], the bulk plasma, sheath, and plasma-sheath transition solutions were treated in both the long mean free path and short mean free path limit. The authors noted that "the plasma-sheath transition is inherently more complicated than either plasma or sheath alone". The exact position of the start of the sheath (breaking of quasineutrality) was somewhat arbitrarily defined by the point ". . . when the Poisson term, neglected in the plasma solution, becomes equal to a *certain fractional part* of either of the other two terms, . . ." Nonetheless, the plasma was clearly separated into two distinct regions: a neutral region far away from the plasma boundary, and a non-neutral region in front of the plasma boundary, called the sheath.

In 1949, Bohm presented the first true criterion for the location where the neutral plasma and the non-neutral sheath meet, now called the sheath edge[8]. (Throughout the remainder this

Determination of the levitation limits of dust particles within the sheath in complex plasma experiments paper, the term sheath edge will be used to refer to the location where quasineutrality is broken.) He determined that ions must enter the sheath with a minimum velocity of $u_+ = v_B \equiv \sqrt{k_B T_e / m_+}$, called the Bohm velocity. However, within the quasineutral bulk plasma, ions flow in the ambipolar electric field, often approximated to be $E \approx -k_B T_e / e\lambda$ where $\lambda$ is the ion mean free path[6]. Therefore, in order to satisfy the Bohm criterion for the sheath edge, ions in the bulk plasma must be accelerated to the Bohm velocity. This realization led to the introduction of the presheath region. The presheath is defined as the quasineutral region in which ions are accelerated so that they obtain the Bohm velocity at the sheath edge. Thus, the plasma is divided into three distinct regions: the quasineutral bulk far away from the plasma boundary, a presheath (also quasineutral), and the sheath region which smoothly connects to the presheath at the sheath edge (see Figure 1a). It is interesting to note that the original derivation of the Bohm criterion neglected ion collisionality and ionization within the sheath, even though two possible requirements for the existence of the presheath are collisionality and ionization within the presheath[9].

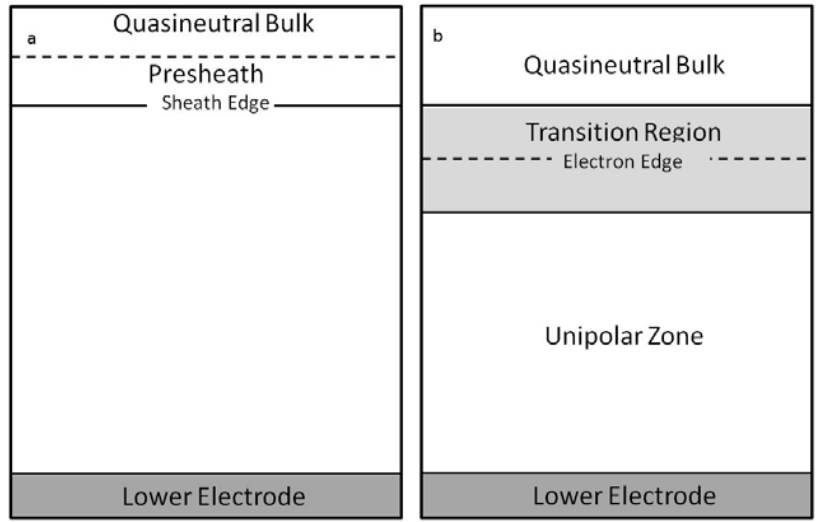

FIG. 1. The regions of a plasma near a plasma boundary such as the lower electrode as described by a) Bohm and b) Brinkmann. The dotted line in (a) represents the boundary between the quasineutral bulk and the presheath and the dotted line in (b) indicates the electron edge.

Although the Bohm criterion has been used for decades to describe the sheath edge, limits to this criterion have been identified. A recent theory by Loizu, et al.[10] shows that the Bohm criterion is not able to accurately describe the sheath edge for systems in which the electron current is the predominant current through surfaces exposed to the plasma. It was found

Determination of the levitation limits of dust particles within the sheath in complex plasma experiments that the sheath edge is located further away from the bounding surface than predicted by the Bohm criterion in these systems. Therefore, in this case, ions can enter the sheath with a velocity less than the Bohm velocity, making the Bohm criterion invalid.

**B. The transition region and the electron edge**

A second model used to describe the sheath is called the step model or electron step, originally introduced by Godyak and Ghanna[11]. The step model assumes that the electron density, $n_e$, is nearly equal to the ion density, $n_i$, far above the electrode and drops instantaneously to zero at the so called electron edge. Brinkmann[12] modified the step model by replacing the sudden drop of the electron density with a transition region where the electron density decays over a width of a few electron Debye lengths. The region below the transition region, where electrons are depleted, is called the unipolar zone and the region above the transition region is called the quasineutral zone since the electron and ion densities are assumed to be equal there (see Figure 1b and Figure 2 in Ref.12 for an illustration). The electron edge, $s$, is the height at which the total electron charge between the electrode surface and $s$ equals the excess positive charge above $s$. This can be expressed as

$$\int_0^s n_e(x)dx = \int_s^{d/2} (n_i(x) - n_e(x))dx, \quad (1)$$

where 0 denotes the electrode surface and $d$ the distance between the electrodes. Note that the electron edge is different from the sheath edge since quasineutrality is broken above the position of the electron edge.

It is also important to note that the transition region is not the same as the above mentioned presheath region, since the presheath, by definition, is quasineutral. Hence $s$ lies approximately one electron Debye length below the presheath region. Therefore, in the classical view, the transition region is in fact part of the sheath. This is illustrated in Figure 1b.

**C. Particle charging: OML theory**

The charging of particles immersed in a plasma environment is most often described by Orbital Motion Limited (OML) theory[13], which describes how plasma particles in an attractive potential well are captured using conservation of energy and angular momentum. By assuming

Determination of the levitation limits of dust particles within the sheath in complex plasma experiments

Maxwellian distribution functions for each plasma species and neglecting ion collisionality and ions in closed orbits[14], the currents collected by a spherical particle (with radius $r_D \ll \lambda_D$, the Debye length) can be readily derived. Within the bulk plasma ($n_e = n_i$) the electron and ion currents to a negatively charged dust particle are then given by

$$I_e = -e\sqrt{8\pi}r_D^2 v_{Te} n_e \exp(\psi) \qquad (2)$$

$$I_i = e\sqrt{8\pi}r_D^2 (1-\gamma\psi) n_i v_{Ti} \qquad (3)$$

where $r_D$ is the dust particle radius, $\psi = e\phi_D/k_B T_e$ is the normalized dust particle surface potential ($\psi < 0$), $v_{Tj} = \sqrt{k_B T_j/m_j}$ is the thermal velocity for species $j$, and $\gamma = T_e/T_i$. To accurately model the current to dust particles within the sheath, the ion current must be modified. This is done through the use of a shifted Maxwellian distribution function which accounts for ion flow within the sheath. The resulting ion current to a negatively charged dust particle within the sheath is now

$$I_i = e\sqrt{2\pi}r_D^2 n_i v_{Ti} \left[ \sqrt{\frac{\pi}{2}} \frac{1+M_+^2-2\gamma\psi}{M_+} erf\left(\frac{M_+}{\sqrt{2}}\right) + \exp\left(\frac{-M_+^2}{2}\right) \right] \qquad (4)$$

where $M_+ = u_+/v_{Ti}$ is the ion thermal Mach number[5]. It should be noted that the ion and electron densities are not equal within the sheath and all plasma parameters are functions of the height above the lower electrode, $z$.

The equilibrium dust particle surface potential, $\phi_D$, is reached when the net current to the dust particle vanishes. The dust particle surface potential is usually related to the dust particle charge through the capacitor model, which for $r_D \ll \lambda_D$ leads to $Q_D = eZ = 4\pi\epsilon_0 r_D \phi_D$, where $Z$ is the dust charge number[15].

**D. The lower instability criterion**

The charged dust particles are levitated by the electric field present within the sheath. Although the electric field extends throughout the sheath, an unstable region has been theorized to exist near the lower electrode in which particles can not be levitated. Previously, Ivlev, et al.[3] derived the condition for this unstable region while investigating the frequency of dust particles oscillating around their equilibrium position including both charge and electric field variation

Determination of the levitation limits of dust particles within the sheath in complex plasma experiments

with height. The real part of the oscillation frequency was found to be

$$(Re\ \omega)^2 \approx -\frac{(Q_D(z)E(z))'_0}{m_D}, \tag{5}$$

where the prime denotes the derivative with respect to $z$ and the subscript 0 denotes the equilibrium position. At heights where $(Q_D(z)E(z))'_0 > 0$ no stable particle position exists [$(Re\ \omega)^2 < 0$]. Therefore, the lower limit of stable dust particle levitation is determined by the local maximum in the electric force.

### III. EXPERIMENTAL PROCEDURES AND NUMERICAL MODEL

The complex plasma experiments discussed here were carried out in the Center for Astrophysics, Space Physics and Engineering Research (CASPER) Gaseous Electronics Conference (GEC) cell[16,17]. The cell contains a grounded, upper electrode and a powered, lower electrode that is capacitively coupled and driven at 13.56 MHz. The electrodes are separated by 2.54 cm. An aluminum plate with a circular cutout 25.4 mm in diameter and 1 mm deep was placed on the lower electrode, creating a horizontal potential which confines the dust. The experiments were performed in Argon plasma at 20 Pa and the amplitude of the RF voltage was varied between 22-66 V. In order to maintain reproducibility, a Kepco external DC power supply was used to maintain a DC bias of -5 V on the lower electrode.

In the first experiment, which will be discussed in Section IV A, the levitation height of particles with different sizes was measured at various RF voltages. Melamine Formaldehyde (MF) particles with a mass density, $\rho$, of 1.510 g/cm$^3$ and diameters of 6.37, 8.89, and 11.93 μm were used. A small number of dust particles (10-50) of a single size were dropped into the plasma and allowed to form a crystal at each RF voltage. Side pictures of the layer were then used to measure the levitation height of the particles.

The second experiment, which will be discussed in Section IV B, was conducted to measure the electric force as a function of height above the lower electrode. MF particles with a diameter of 8.89 μm were dropped into the plasma, and their trajectories were recorded using a Photron 1024 high-speed camera at 1000 frames per second. Particle trajectories were then used to reconstruct the electric force on the particle as a function of height.

Finally, a self-consistent fluid model was employed to model the plasma parameters. A complete description of the fluid model used can be found in the literature[17]. The plasma

Determination of the levitation limits of dust particles within the sheath in complex plasma experiments parameter profiles as a function of height above the lower electrode were calculated for the geometry of the CASPER GEC cell. Using these profiles in the current balance equation that includes ion drift, the dust potential, $\phi_D(z)$, was obtained. The dust charge as a function of height was then found from the dust potential through use of the capacitor model. For comparison with experiment, the pressure was fixed at 20 Pa, and the driving potential amplitudes, $V_{RF}$, were varied. In Section IV C, these fluid model results are presented along the symmetry axis of the discharge chamber where $z = 0$ represents the lower electrode.

## IV. RESULTS

### A. Experimental levitation of multiple particle sizes

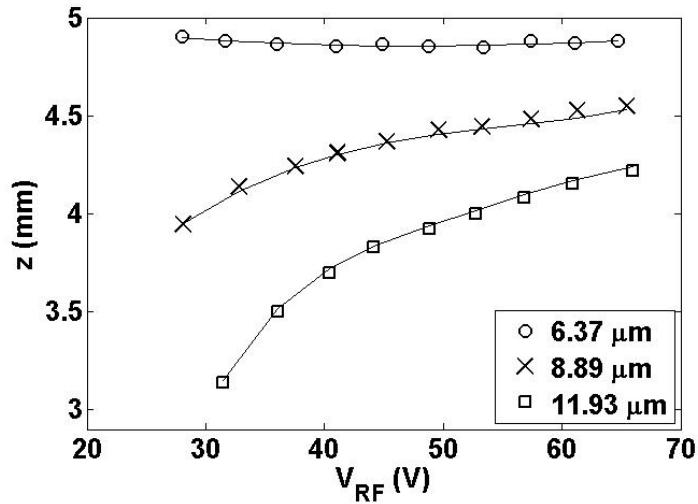

FIG. 2. Experimentally measured levitation heights of different-sized particle layers as a function of RF voltage at a pressure of 20 Pa. The levitation height of each particle size was measured in separate experiments. Lines are added to guide the eye.

Figure 2 shows the height above the lower electrode of single dust layers for three different dust sizes as a function of RF voltage. At low RF voltages (≈ 30−45V ) the levitation heights of the 8.89 μm and 11.93 μm particles increase rapidly with increasing RF voltage, but at higher RF voltages (> 45V ) the particle height approaches a constant value. In contrast, the 6.37 μm particles do not appear to be affected by the changing RF voltage across the range of voltages tested.

Determination of the levitation limits of dust particles within the sheath in complex plasma experiments

The particles are assumed to levitate at the height where the gravitational force is equal to the electric force on the charged particle. Since the gravitational force is constant for a specific particle size, the levitation height dependence on RF voltage must be due to changes in the electric force. Therefore, the electric force as a function of height above the lower electrode is required to fully understand the behavior shown in Figure 2.

**B. Experimental determination of the electric force**

A second experiment was used to measure the electric force, $F_E$, as a function of height above the lower electrode. In this experiment, the trajectories of 8.89 μm particles were recorded as they fell from above the upper electrode to their levitation height. For this case, the vertical motion of a particle is described by

$$m_D \ddot{z} = F_E(z) - m_D g - m_D \beta \dot{z} \tag{6}$$

where $\beta$ is the Epstein drag coefficient[18], $m_D$ is the dust particle mass, and $z$ is the height of the particle above the lower electrode. $\beta$ was calculated to be 30.5 s$^{-1}$ using of the equation

$$\beta = \frac{8}{\pi} \frac{P}{\rho r_D v_{th,n}} \tag{7}$$

where $P$ is the gas pressure, $\rho$ and $r_D$ are the mass density and radius of the dust particle, respectively, and $v_{th,n}$ is the thermal velocity of the neutral gas. Using the theoretical value for $\beta$ and the trajectory data of the particles, the electric force, $F_E(z)$, was calculated and is shown in Figure 3. The horizontal line indicates the magnitude of the gravitational force acting on the particle.

The levitation height is again assumed to be the height where the gravitational force and electric force are equal. Therefore, the levitation height dependence on RF voltage found in this experiment confirms that found in the previous experiment (Section IV A) (i.e., increasing RF voltage increases equilibrium height of the particles, but this dependence is nonlinear with RF voltage). The electric force for $z < 5$ mm increases as RF voltage increases as expected, but for $z > 5$ mm, the electric force curves coincide, indicating that the electric force is independent of RF voltage in this region within the errors of the experiment. Comparing this with the previous experiment (Section IV A) it is evident that the 6.37 μm particles levitate near the

Determination of the levitation limits of dust particles within the sheath in complex plasma experiments height where the electric force curves begins to coincide, while the larger particles (8.89 μm and 11.93 μm) levitate below this point.

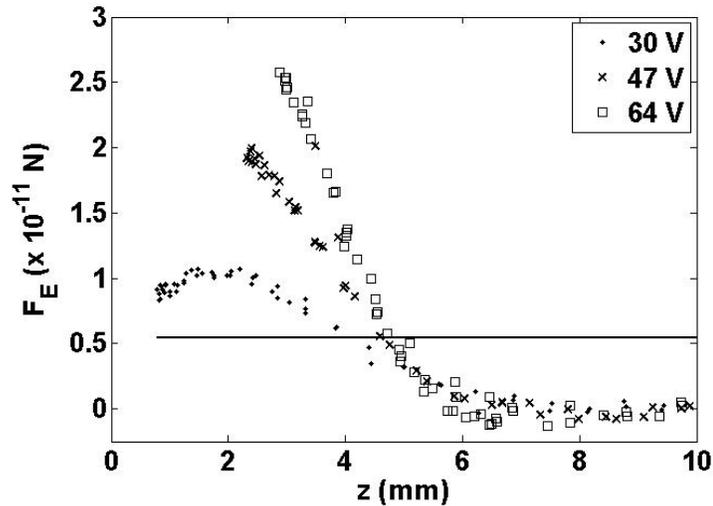

FIG. 3. The experimentally determined electric force profile at a pressure of 20 Pa for various RF voltages. The solid, horizontal line indicates the magnitude of the gravitational force acting on an 8.89 μm particle.

Therefore, the convergence of the electric force is responsible for the constant levitation height of the 6.37 μm particles. While this explains the behavior of the particles shown in Figure 2, knowledge of the plasma parameters that determine the electric force profile is required to understand why the electric force curves coincide as they do. As previously stated, plasma parameters can not be easily determined experimentally, so a fluid model was employed.

## C. Fluid Model Results

A fluid model was used to calculate $n_i, n_e, v_{Te}, v_{Ti}, T_e$, and $M_+$ as a function of height employing the same plasma conditions and geometry used in the experiments. The density ratio, $\alpha(z) = n_i(z)/n_e(z)$, found from the fluid model is shown in Figure 4. The dust surface potential is found by inserting the plasma parameters found from the fluid model into the electron and ion current equations (equations 2 and 4) and determining the value of $\phi_D(z)$ at which the electron and ion currents are equal. The dust surface potential profiles obtained in this manner are shown in Figure 5. Once $\phi_D(z)$ is known, the dust particle charge can be calculated through use of the capacitor model. Combining the dust charge with the fluid model solution for the electric field, $E(z)$, shown in Figure 6, the electric force is found as $F_E(z) = Q_D(z)E(z)$.

Determination of the levitation limits of dust particles within the sheath in complex plasma experiments

Figure 7 shows the resultant electric force for 8.89 μm and 16 μm particles. The solid, horizontal lines indicate the gravitational force for each particle size.

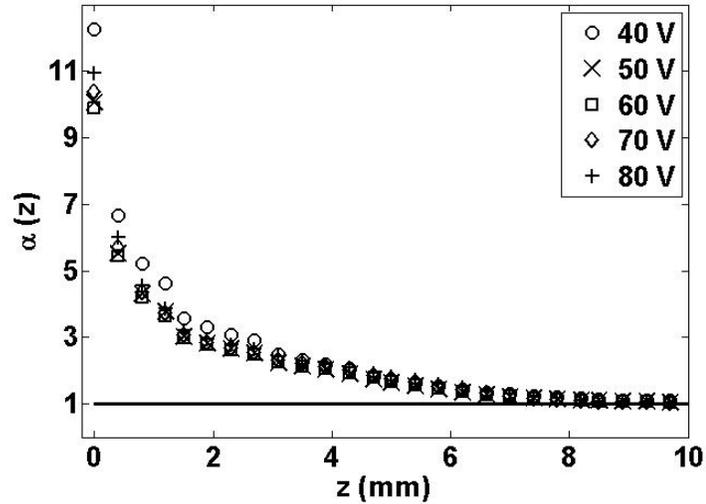

FIG. 4. The ratio of the ion density to the electron density, $\alpha(z)$, in the sheath for various driving potentials, $V_{RF}$, at 20 Pa obtained from the fluid model. The horizontal line indicates where the electron and ion densities are equal.

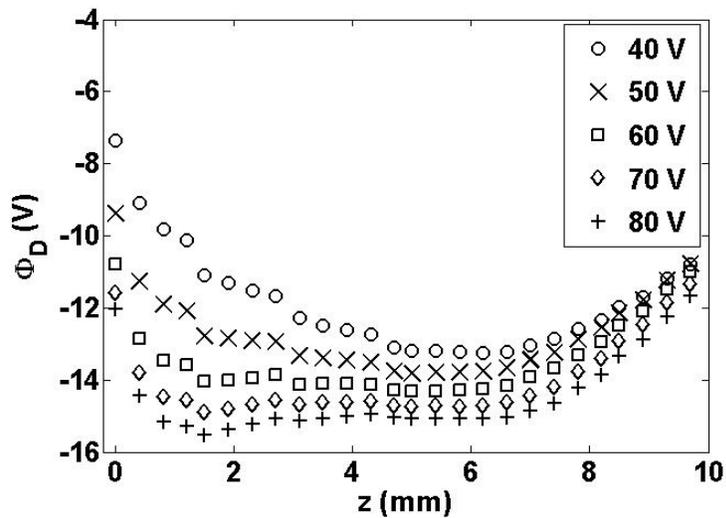

FIG. 5. The dust potential for various driving potentials, $V_{RF}$, at a pressure of 20 Pa obtained from the fluid model.

Again, the equilibrium levitation height of the dust particles is assumed to be the height at which the electric and gravitational force are equal. The electric force profile found from the fluid model displays the same trend as that found in the experiment (Figure 3) in that the electric force reaches a maximum value near the lower electrode and decreases with increasing height

Determination of the levitation limits of dust particles within the sheath in complex plasma experiments above that point. Figure 8 shows the levitation heights for various particle sizes and RF voltages obtained from Figure 7. As observed in the experiment, larger particles show an increase in particle levitation height with RF voltage, while the levitation height for the smaller particles appears to be nearly independent of the RF voltage.

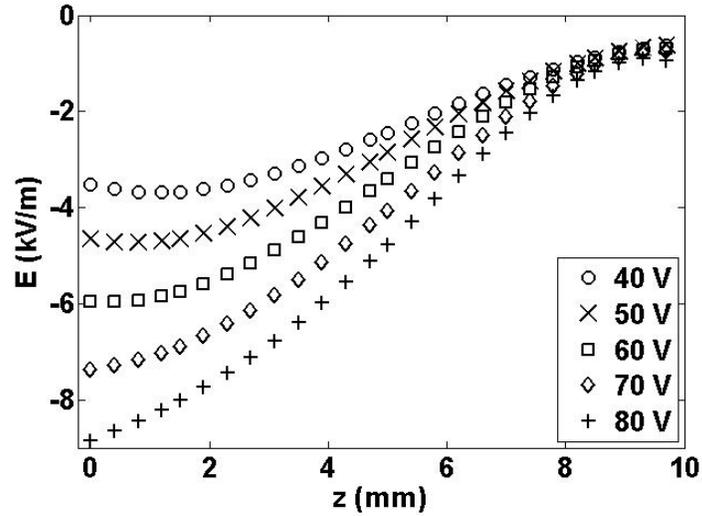

FIG. 6. The electric field profile for various RF voltages at 20 Pa obtained from the fluid model.

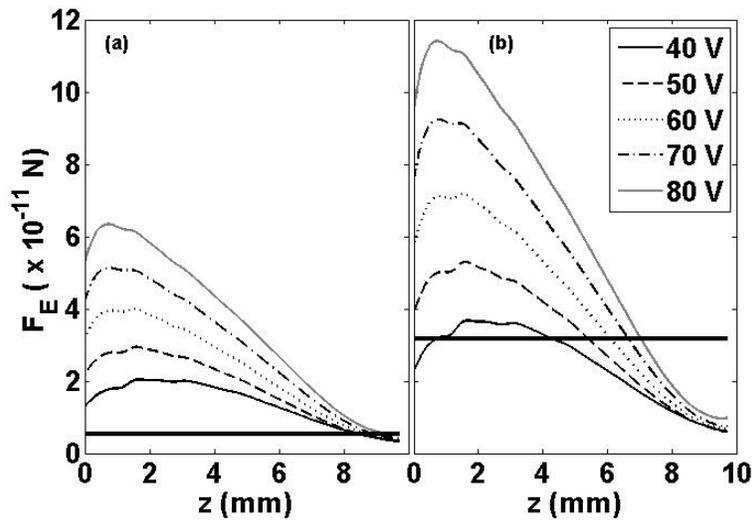

FIG. 7. The electric force profiles at various RF voltages and a pressure of 20 Pa for (a) 8.89 μm and (b) 16 μm diameter particles obtained from the fluid model. The solid horizontal line indicates the magnitude of the gravitational force acting on each particle.

Determination of the levitation limits of dust particles within the sheath in complex plasma experiments

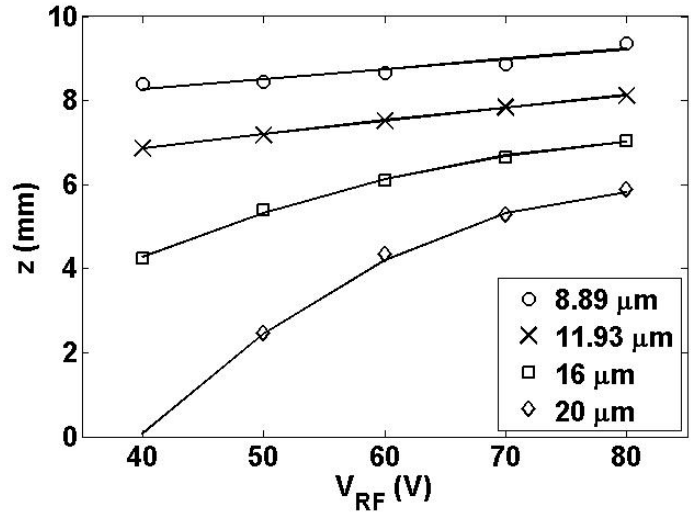

FIG. 8. Levitation heights of different-sized particles as a function of RF voltage at 20 Pa, obtained from the fluid model. Solid lines are added to guide the eye.

## V. DISCUSSION

### A. Comparison of experiments and fluid model

Comparing figures 2 and 3 to figures 8 and 7, respectively, it is evident that the fluid model employed gives representative results which can be compared to those in the experiment. Although the magnitude of the gradient of the electric force found from the fluid model is much smaller than that found in the experiment, the maximum electric forces found only differ by approximately 17% and the heights at which the maximum force occurs are within 0.5 mm of each other. While the fluid model results are not identical with the experimental results, both exhibit the same physical behavior, allowing the plasma parameter profiles obtained from the fluid model to be used to explain the behavior observed in the experiments. With these results, the location of the lower levitation limit, which marks the top of the unstable region, and the sheath edge, as illustrated by Figure 9, will be determined.

In both the model and experiment it was found that particles with $dz/dV_{RF} \approx 0$ (the 6.37 μm particles in the experiment and 8.89 μm particles in the model) are levitated at a height where the electric force profiles for all RF voltages converge. The fluid model was used to determine the physical reason for this convergence. The electric force profiles found with the fluid model converge at $z \approx 9$ mm, which is the same height that convergence is observed in the

Determination of the levitation limits of dust particles within the sheath in complex plasma experiments electric field, dust surface potential, and density ratio. The density ratio, shown in Figure 4, shows that at this height the electron and ion densities begin to diverge, marking the edge of the quasineutral plasma - the sheath edge. It is important to note that this does not always imply that particles with levitation heights which show $dz/dV_{RF} \approx 0$ are located at the sheath edge. In some situations, the ambipolar electric force within the presheath or bulk may be sufficient to levitate particles with a sufficiently small radius. The levitation height of particles located within the quasineutral plasma will also display levitation heights with $dz/dV_{RF} \approx 0$. Therefore, the largest particles to display levitation heights with $dz/dV_{RF} \approx 0$ can be used to locate the sheath edge. While this method for determining the location of the sheath edge is easily used with the fluid model, experimental determination of the sheath edge is less precise. Such experiments are limited by the sizes of manufactured dust available, resolution of the cameras, and other sources of error due to the precision of the equipment used. Therefore, particles that maintain a nearly constant height across a wide range of RF voltages are at or near the sheath edge, but an infinite number of particle sizes and resolution are required to precisely locate the sheath edge.

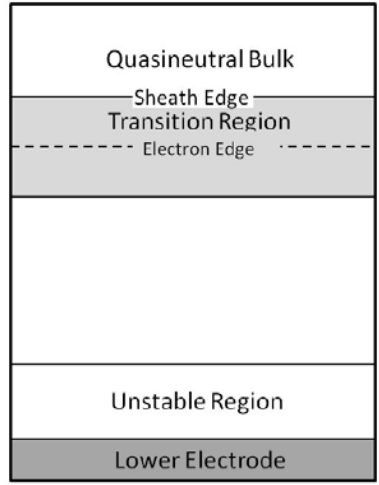

FIG. 9. The regions of a plasma near the plasma boundary discussed in this paper. The dotted line indicates the electron edge.

It is also important to investigate the electric field that particles located at or near the sheath edge are levitated in. The fluid model electric field curves shown in Figure 6 are best fit by a third-order polynomial. Near the center of the sheath, the electric field is found to be approximately linear, transitioning to a nearly constant value near the sheath edge. Therefore, analyses that model the electric field as a linear function of height across the entire sheath may

Determination of the levitation limits of dust particles within the sheath in complex plasma experiments

under- or overestimate the magnitude of the electric field at a given height depending on the approximation used. To experimentally determine locations where the electric field deviates from this approximately linear region, the levitation heights of particles should be measured across a wide range of RF voltages. Particles having levitation heights independent of RF voltage (such as the 6.37 μm particles in the experiment described above) are located outside the linearly increasing portion of the electric field.

Finally, as discussed in Section II D, an unstable region has been theorized to exist near the lower electrode where $(Q_D(z)E(z))'_0 > 0$. Figure 7 shows that the electric force does in fact have a positive gradient near the lower electrode, indicating an unstable region where particles cannot be levitated. This region extends from the lower electrode to the height where the electric force attains a maximum value. The 16 μm particles in Figure 7b, show two levitation heights for the 40 V curve. These particles will only levitate at the upper equilibrium height since the lower equilibrium height is located within the unstable region, making it an unstable equilibrium point. This fact was recently used in an experiment to manipulate the number of particles in a vertically extended dust chain[19]. Figure 10 shows the minimum levitation height versus RF voltage as

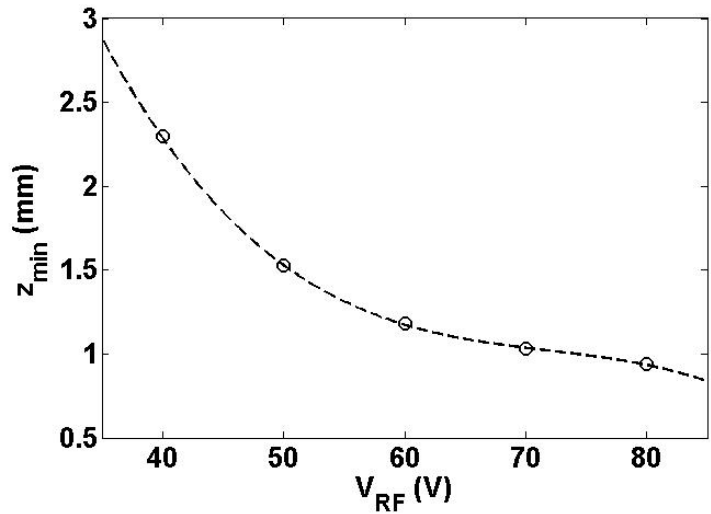

FIG. 10. The minimum levitation height for dust particles as a function of RF voltage as obtained in the fluid model. The dashed line is added to guide the eye.

obtained by finding the best-fit polynomial to the electric force profiles from Figure 7 and locating the heights at which the derivative of the polynomial was zero, indicating the location of the maximum electric force. Larger RF voltages create a stronger electric field near the lower

Determination of the levitation limits of dust particles within the sheath in complex plasma experiments electrode, in turn allowing the dust to levitate at lower heights before entering the unstable region and subsequently falling to the lower electrode.

It is interesting to note that the location of the minimum levitation height is independent of the dust particle size and is therefore a property of the system itself. This can be seen from Figure 7 which shows that the maximum in the electric force for a specific RF voltage occurs as the same height for both the 8.89 μm and 16 μm particles, only the magnitude is different. The electric force does linearly depend on the dust particle radius (through the dust charge), but this only serves to change the magnitude of the electric force and not the location of the maximum. All other parameters used to calculate the electric force ($E(z), \phi_D(z)$, etc.) only depend on the system parameters and, therefore, do not affect the location of the maximum.

From the experimental results shown in Figure 3, we see that the minimum levitation height of the particles is approximately 2 mm at 30 $V_{RF}$. The minimum levitation height (or maximum in the electric force) for higher RF voltages could not be accurately determined experimentally at these settings. However, the validity of this theory can be demonstrated through use of another experimental setup. The 8.89 μm dust particles were dropped into a second CASPER GEC cell, which has a smaller electrode spacing and was operated at a pressure of 27.7 Pa. The resultant electric force profiles shown in Figure 11 qualitatively agree with that of Figure 3 and additionally show a maximum of the electric force near the lower electrode.

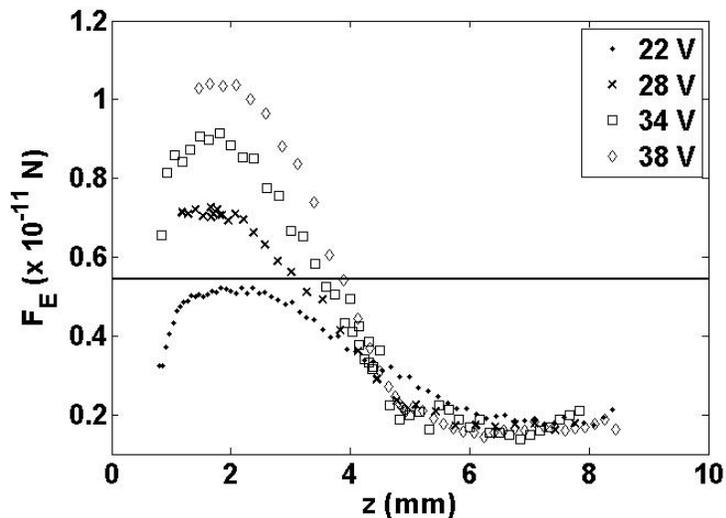

FIG. 11. The electric force profiles for an 8.89 μm dust particle in the second CASPER GEC cell at a pressure of 27.7 Pa. The solid line indicates the gravitational force on the particle.

Determination of the levitation limits of dust particles within the sheath in complex plasma experiments

## B. Comparison with current sheath models

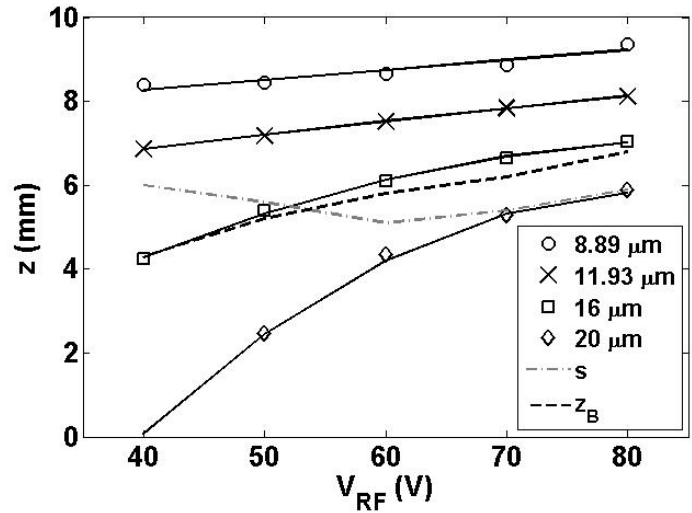

FIG. 12. Levitation heights of different-sized particles as a function of RF voltage at 20 Pa, obtained with the fluid model. Solid lines are added to guide the eye. Dashed lines indicate the Bohm point and electron edge.

In Figure 12, the dust levitation height is compared to previously published standard sheath models. Increased RF voltage causes ions to attain the Bohm velocity at a greater height above the powered electrode. As a result, the height at which ions reach the Bohm velocity, $z_B$, (as determined from the fluid model results) increases with RF voltage. The 16 μm particles levitate at or slightly above the Bohm point for the voltages shown. Therefore, both the 8.89 and 11.93 μm particles are located outside the classically defined Bohm sheath and that the Bohm criterion does not correlate with the location of the sheath edge in this case.

Applying the definition of the electron edge as described by Brinkmann to the fluid model results, the electron edge is found to be located between 5.1 mm and 6.0 mm for the RF voltages tested. The results are also shown in Figure 12. Assuming that the transition region is centered at $s$ and has a width of a few electron Debye lengths ($\lambda_{De} \approx 2$ mm), then both the 8.89 μm and 11.93 μm particles are found to be located above the electron edge but still within the transition region. The 8.89 μm particles are located near the top of the transition region, very close to the height where quasineutrality is obtained, again confirming that they mark the location of the sheath edge.

## VI. CONCLUSIONS

A combination of experiment and numerical modeling was used to determine the levitation heights of dust particles in a GEC RF reference cell. The methods described in Section IV lead to a new method to determine the location of the edge of the sheath above the lower electrode. In the first experiment, dust particle levitation heights for various RF voltages and dust particle sizes were investigated. It was experimentally determined that while the levitation height of small particles (6.37 μm) were unaffected by the RF voltage, larger particles (8.89 and 11.93 μm) exhibited a strong dependence on RF voltage (Figure 2). In an effort to explain this observed behavior, a second experiment was performed, in which particles were dropped into the plasma from above and their trajectories used to calculate the electric force profile shown in Figure 3. While the electric force profile confirmed the dependence of particle levitation height on RF voltage, as observed in the first experiment, this dependence could not be fully understood without knowledge of the plasma parameters, such as $n_e$ and $n_i$, particle charge, and the electric field as a function of height. A fluid model was employed to obtain these plasma parameter profiles, and hence to determine the dust surface potential (Figure 5) and electric field (Figure 6) as a function of $z$ for various driving potentials. This in turn, allowed the electric force profile to be determined as a function of RF voltage (Figure 7). The electric force profile was then used to find the levitation heights for various dust sizes (Figure 8). Fluid model results were found to be in good agreement with the experiments and therefore, used to explain the experimentally observed behavior seen in Figure 2.

These results lead to a new, *in situ* method to determine the approximate location of the sheath edge, defined as the point where quasineutrality is broken. The minimum height at which particles display $dz/dV_{RF} \approx 0$ across a wide range of RF voltages theoretically locates the boundary between the transition region and quasineutral plasma (the sheath edge) but due to the limited resolution of the experiment, particles that display this behavior can only be said to be near the sheath edge. As the sheath edge locates the upper boundary of the transition region, an approximate location of the electron edge (assuming the electron edge is a few Debye lengths below the top of the transition region) can also be found. Particles located at or near the sheath edge are levitated outside the approximately linearly increasing portion of the electric field,

Determination of the levitation limits of dust particles within the sheath in complex plasma experiments indicating that it is incorrect to model the electric field thoughout the entire sheath as linear, as is common practice.

Finally, we note that there is a minimum particle levitation height for each RF voltage. This height coincides with the maximum value of the electric force. This height can be experimentally found using the dropping technique discussed in Section IV B or through the use of other experimental methods that change the levitation height of the particle without changing the plasma parameters, such as the hypergravity experiment[20].

While the fluid model is able to qualitatively reproduce the results found in the experiments, quantitative differences between the two are likely due to approximations and limitations of the model. The fluid model does not include non-linear electron heating and secondary electron emission from the electrode. These terms can be important at low pressures, and inclusion of these terms would result in a more negatively charged dust particle. Ion collisionality and trapped ions, effects which are significant at higher pressures, were also not included. These terms tend to increase the ion flow to a particle, making the dust particle more positively charged. While inclusion of these effects is important, we believe these effects are secondary and will not significantly alter our results. Finally, the drift-diffusion approximation was used to approximate particle fluxes in the fluid model. While conditions for this approximation are easily fulfilled for electrons, Argon ions may not always have a sufficient collision frequency to justify the use of this approximation.

## ACKNOWLEDGMENTS

This material is based upon work supported by the National Science Foundation under Grant No. 0847127 and by the Texas Space Grant Consortium through a graduate fellowship.